\documentclass[pre,twocolumn,showpacs,preprintnumbers,amsmath,amssymb]{revtex4} 
\usepackage{amsmath}
\usepackage{graphicx}
\usepackage{dcolumn}
\usepackage{bm}

\def\max{\mathrm{max}}
\def\lmax{l_{\mathrm{max}}}
\def\d{\mathrm{d}}

\begin{document}
\title{Gompertz mortality law and scaling behaviour of the Penna model}
\author{J.~B.~Coe$^{1,*}$ and Y.~Mao$^{2}$}
\affiliation{$^1$Biomathematics \& Statistics Scotland\\
James Clerk Maxwell Building, The King's Building, Edinburgh, EH9 3JZ
$^2$School of Physics and Astronomy, University of Nottingham, University Park, Nottingham NG7 2RD, United Kingdom}
\thanks{ J.~B.~C.~is also affiliated to the Institute of Evolutionary Biology, University of Edinburgh and the Department of Physics and Astronomy, University of Edinburgh.}


\begin{abstract}
The Penna model is a model of evolutionary ageing through mutation
accumulation where traditionally time and the age of an organism are
treated as discrete variables and an organism's genome by a binary bit
string.  We reformulate the asexual Penna model and show that a universal
scale invariance emerges as we increase the number of discrete genome
bits to the limit of a continuum. The continuum model, introduced by Almeida and Thomas in [Int.~J.~Mod.~Phys.~C, {\bf 11}, 1209 (2000)]
can be recovered from the discrete model in the limit of infinite bits coupled with a
vanishing mutation rate per bit. Finally, we show that scale
invariant properties may lead to the ubiquitous Gompertz Law for
mortality rates for early ages, which is generally regarded as being
empirical.
\end{abstract}

\maketitle

\section{Introduction}
The Penna model was devised in 1995 by T.~J.~P.~Penna \cite{Penna} to
model the process of evolutionary ageing through mutation
accumulation. The idea that natural selection would permit behaviour
such as ageing is initially baffling: it would seem that survival of
the fittest would remove any such detrimental behaviour. Medawar
proposed \cite{medawarAgeing} that certain genes may be age specific
in their effects; if such genes are harmful and are activated late on
in the reproductive life of an organism, natural selection against
them will be much weaker than if they had become active earlier in the
organism's life. Given the existence of such genes, it can be
anticipated that harmful genetic conditions will become more common as
an organism ages giving rise to increasing mortality rates with
age. The Penna model is a means to model the evolution of an
age-structured population under the influence of age-specific harmful
mutations \cite{stauffer}.

Traditionally, mortality rates are known to rise exponentially for
early ages, giving rise to the Gompertz law \cite{gompertzLaw} of
mortality.  More recent experiments using much larger sample
populations have shown that the mortality rate for advanced ages is
seen to slow substantially giving rise to a mortality plateau or peaks
\cite{mortPlat1,mortPlat2,mortPlat3,mortPlat4}. It has been shown that
a modified Penna model while continuing to show Gompertz growth in
mortality rates at early ages can also exhibit a mortality plateau at
advanced ages \cite{Coe_Mao_Cates}.

The original Penna model is discrete in nature, with time represented
by an integer and an organism's genome by a bit-string. 
Each $0$ on the bit-string represents a healthy site; each $1$ is a harmful
mutation which becomes active once the organism reaches age $x$ where $x$ is the 
index of the site on the bit-string. Having activated $T$ harmful mutations
an organism dies.
The bit-string is taken to be finite in length (usually 32 bits) and each newborn
organism has a number of mutations $M$ introduced into the bit-string. These mutations
are taken to be harmful so can only turn healthy sites into unhealthy ones - a mutation
on an unhealthy site is ignored. This assumption is relaxed in \cite{posMut} where
a small rate of positive mutation is allowed: we confine ourselves here to the case of only
harmful mutations.

Scaling behaviour was considered by Malarz \cite{stringlength} who investigated the
effects of different bit string lengths on the Penna model. Malarz inquired as to whether
large bit strings were required or whether one could expect, after appropriate scaling of other
parameters, one would get the same results for different genome lengths. Investigating the effects
of string length through simulation, Malarz was unable to find scaling in the Penna model.

Almeida et al.~\cite{AlmeidaContPenna} later considered a continuous Penna model and
for certain mutation regimes were able to find simple scaling relations. To obtain such scaling
the authors decoupled the string-length and mutation rate so that the probability of finding
a given number of mutations in a given string length was given by a poisson distribution.
They also observed that the Penna model is able to sustain a maximum possible lifespan in steady-state,
which we call $\lmax$: if the imposed string length is greater than $\lmax$ then it will have no effect on
the properties of the population; if it is less than $\lmax$ then the imposed string length will
impose a maximum lifespan on the population and the distribution will be accordingly altered.
The authors suggested that the size of timesteps in a discrete Penna model may have an effect on scaling behaviour
but did not investigate the size or nature of this effect.

Brigatti et al.~\cite{ref2} investigated scaling in a sexual Penna model through simulation and 
suggested that results from the continuum model of Almeida et al.~\cite{AlmeidaContPenna} were
not readily mapped onto the discrete model employed in simulation. Scaling effects in the sexual model were
also investigated by Laszkiewicz et al.~\cite{Laszkiewicz} through simulation.


In this paper we extend our previous analytical solution of the asexual model
\cite{Coe_Mao_Cates,Analytical_solution} to examine the scaling
behaviour. We show that the scale invariance emerges as we increase
the number of discrete genome bits, and that the scaling becomes exact
in the continuum case, which can be regarded as the limit of infinite
genome bits coupled with a vanishing mutation rate per bit. 
This establishes a clear relationship between the distribution, parameters and scaling behaviour
of the continuum model of Almeida et al.~and those of the traditional discrete model.
Finally, we use scale invariance to analytically show that at early ages
mortality rates grow exponentially in accordance with the
Gompertz law \cite{gompertzLaw} which, for the lack
of a general proof, is still generally regarded as being empirical
\cite{rose}.

\section{A continuous Penna model}
The asexual Penna model can be reformulated
\cite{AlmeidaContPenna,Analytical_solution} so that rather than
considering discrete timesteps, time is treated as a continuous
variable, $t$. The bit-string of an organism is replaced by an axis
representing the genome: position $x$ on the genome is examined at age
$x$. Harmful mutations are then represented by $\delta$ functions along
the genome. After accumulating $T$ $\delta$ functions an organism dies.

For our analytical solutions, we concern ourselves primarily with
$T=1$ as generalizing a $T=1$ solution for a continuous model will be
no more difficult than generalizing a discrete $T=1$ model, as done
previously \cite{Coe_Mao_Cates}.  In the continuum Penna model an
organism reproduces at a constant rate $b$ and dies at age $x$ where
its genome has its first harmful mutation ($\delta$ function) at position
$x$.

An organism can be characterised by its age $x$ and its genetic
lifespan $l$ (the position on the genome of the first harmful
mutation). Neither $x$ nor $l$ are constrained to be integers.
$n(x,l)$ is now a density of organisms so that the number of organisms
with age and genetic lifespan in the range $x\to x+\d x$ and $l\to
l+\d l$ is given by $n(x,l)\d x \d l$.  The probability of giving
birth in time $\d t$ is given by $b \d t$, the probability of a
mutation being introduced in length $\d l$ is given by $\beta \d l$.
These definitions are consistent with the discrete Penna model where
sites can be interpreted as infinitessimal lengths of genome and
timesteps as infinitessimal units of time.  The probability of no
mutations occuring in length $\d l$ is $1-\beta \d l$ which is
$e^{-\beta \d l}$ for infinitessimal $\d l$.

\begin{figure}
\begin{center}
 \includegraphics[width=3in]{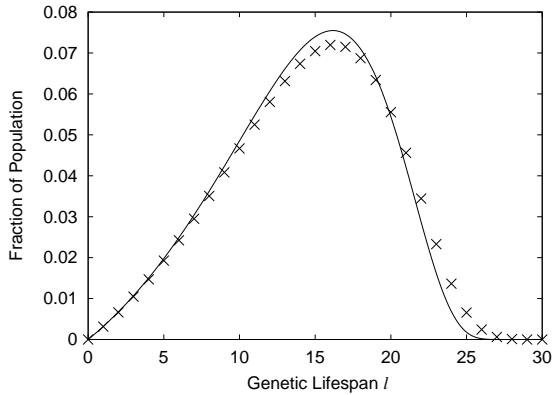}
 \caption{\label{figContPM1}A plot of genetic lifespan distribution for a discrete ($\times$) and continuous Penna model with $l_{\mathrm{max}}=30$.}
\end{center}
\end{figure}

Newborn organisms may be produced as unmutated copies of organisms
with equal genetic lifespan, or as mutated copies of naturally longer
lived organisms.  An equation can then be constructed for the
production of new organisms within the population for the
infinitesimal time period of $t$ to $t'=t+\d t$
\begin{eqnarray}
&&n(0,l)_{t'}\d t\,\d l = b \d t \d l\; e^{-\beta l} \int_{0}^{\infty} \d x\, n(x,l)_t \nonumber\\
&+& b \d t\, \beta \d l \;e^{-\beta l} \int_{0}^{\infty}\d x \int_l^\infty \d l' n(x,l')_t
\end{eqnarray}
where subscripts $t$ and $t'$ denote time.
At steady-state, the subscripts may be dropped, the above equation can
be simplified and an 
expression obtained \cite{Coe_Mao_Cates,Analytical_solution} for the
relative sizes of population densities (see Fig.~\ref{figContPM1}).
\begin{eqnarray}
\frac{n(l+x)}{n(l)}&=&\frac{l+x}{l}\frac{e^{\beta l}-bl}{e^{\beta(l+x)}-b(l+x)}\nonumber\\
&\times&\exp \left[ {\int_{l}^{l+x}\frac{\beta bl'}
{bl'-e^{\beta l'}}\d l'} \right].
\end{eqnarray}

For a steady-state to exist there must be a longest lived
sub-population which is self-sustaining, {\it i.e.} not reliant on
mutated births.  No other sub-population can be self-sustaining if the
population is to remain bounded, as shorter-lived organisms can always
be created by mutated copies of longer-lived ones. For the
longest-lived sub-population to be self-sustaining, each organism must
produce one perfect copy of itself during its lifetime
\begin{eqnarray}
l_{\mathrm{max}}be^{-\beta l_{\mathrm{max}}}=1.
\end{eqnarray}
All other populations, with $l<l_{max}$, gain from mutated births of
the longest lived, so unmutated birth per individual must, on average,
be less than unity 
\begin{equation}
l \; be^{-\beta l}<1 \quad
\quad  \forall \quad l<l_{max}.
\label{eq:trouble}
\end{equation}
These conditions can be combined to give \cite{Analytical_solution}
\begin{eqnarray}
l_{\mathrm{max}}&\leq&\frac{1}{\beta}\\
b&=&\frac{1}{l_{\mathrm{max}}}e^{\beta l_{\mathrm{max}}}.
\label{eq:trouble2}
\end{eqnarray}
In the discrete Penna model the probability of no mutation for $1$
site or bit is $1-m$ where $m$ is the mutation rate per site. The
probability for $l$ sites without mutations would be $(1-m)^{l}$.  In
the limit of $m \to 0$, $(1-m)^l \approx e^{-ml}$ and therefore $m$
play the same role as $\beta$ in the continuous case, where the
probability of no mutations in genome length $l$ is $e^{-\beta l}$.
Thus we can identify the continuum Penna model as the limit of the
discrete model as the mutation rate per site tends to zero.  For
vanishingly small units of discretization, a discrete model becomes a
continuous one. A measure of the extent of discretization is the size
of one of the discrete units divided by the total size of the system;
for the Penna model this is $\frac{1}{l_\max}$. As the extent of
discretization gets smaller, $l_\max$ tends to infinity which implies
a vanishing mutation rate. Thus, the two limits of mutation rate
tending to zero, and of increasingly fine grained discretization, are
identical.

\section{Scaling properties}

We examine the discrete and continuous Penna models in turn to examine how they behave under rescaling. Informed
by this behaviour we interpret the continuous model as the limit of a discrete model with a vanishingly small
mutation rate.

\subsection{The discrete Penna model}

The traditional asexual Penna model has one unit of discretization for each unit of time.
It is possible to rescale the discrete Penna model so that each unit of time is broken up into several timesteps.
This can be done by taking a Penna model with a maximum lifespan of $a l_\max$ and rescaling $l$ so that, in
the rescaled time units, the model has maximum lifespan $l_\max$ and $a$ distinct timesteps in one unit of time.
For example, an $l_\max=30$ model could be rescaled to give an $l_\max=15$ model with two timesteps per unit of time.

When discussing rescaled Penna models we require that the steady-state conditions
are invariant under rescaling. For a population with $l_\max=60$, the steady-state conditions should be the
same regardless of how many timesteps one unit of time has been broken up into. 
For steady-state conditions to be invariant under rescaling the population with genetic lifespan $l_\max$ must
be self-sustaining and all other populations partly dependent on mutation. The first condition can be
written as:
\begin{equation}
bl_\max e^{-\beta l_\max}=1
\end{equation}
A model rescaled by a factor $a$ will allow $n(l_\max-\frac{1}{a})$ to
exist, where $a$ gives the number of units of discretization per time
interval.  The same conditions, equation
(\ref{eq:trouble}-\ref{eq:trouble2}), apply as before.
A population is then identified by the largest (unscaled) value of $l_\max$ it can sustain. When steady-state 
is required to be robust under rescaling of the model a population can be uniquely identified by the maximum
genetic lifespan it can maintain.

For rescaled models to be the same they should give the same
population sizes at comparable points up to an arbitrary scaling
factor.  If the discrete Penna model is scale invariant, it should be
possible to rescale a model to obtain an
unscaled model with shorter $\lmax$. For instance: an $\lmax=30$ model
with scaled by a factor of $2$ will have a rescaled maximum lifespan of
$15$; if the Penna model is scale invariant this rescaled model will, at comparable points
give identical results to an unscaled $\lmax=15$ model (up to a constant
normalisation factor for finite size scaling). Where $n_{30}$ denotes
a model with unscaled maximum lifespan $30$, we require that
$n_{30}(2l) / n_{15}(l)$ is constant. In a general case for models to
be identical after scaling we require that
\begin{equation}
n_{a\lmax}(al)
\propto n_{\lmax}(l).
\end{equation}
This can be satisfied, eliminating the constant of proportionality by
\begin{equation}
  \frac{n_{a\lmax}(al+a)}{n_{a\lmax}(al)}
=\frac{n_{\lmax}(l+1)}{n_{\lmax}(l)}.
\end{equation}
In the case of $a=2$ we require that
\begin{equation}
\frac{n_{[l_\max,2]}(l+1)}{n_{[l_\max,2]}(l+\frac{1}{2})}\frac{n_{[l_\max,2]}(l+\frac{1}{2})}{n_{[l_\max,2]}(l)}
=\frac{n_{[l_\max,1]}(l+1)}{n_{[l_\max,1]}(l)}.
\end{equation}

\begin{figure}
\begin{center}
 \includegraphics[width=3in]{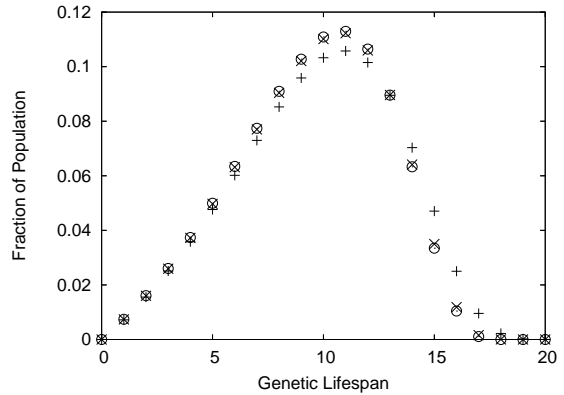}
 \caption{\label{figScalingTrio}A plot of genetic lifespan distribution for 
an unscaled Penna model with $\lmax=20$ ($+$), a model with $\lmax=200$ scaled down by a factor of $10$ ($\times$) and a model with $\lmax=2000$ scaled down by a factor of $100$ ($\circ$). Only comparable points have been shown.}
\end{center}
\end{figure}

For a Penna model to have $l_\max$ a factor of $a$ greater,
the mutation rate and birth rate must be a factor of $a$ smaller. If the parameters of the model
which is rescaled are labelled as $l'$, $\beta'$, $m'$ and $b'$; then scaled and unscaled parameters
are related by:
\begin{eqnarray}
l'&=&al\\
l_\max'&=&a l_\max\\
\beta'&=&\frac{\beta}{a}\\
b'&=&\frac{b}{a}.
\end{eqnarray}

Application of these scaling rules; the recursion relation between successive sub-populations at steady-state;
and our condition for scale invariance of the model gives a relation, in terms of
birth and mutation rate, which must be satisfied for the discrete model to be scale invariant.

For a rescaling by a factor of $2$ we require that
\begin{eqnarray}
&&\frac{ e^{\beta l} - bl }{ e^{\beta(l+\frac{1}{2})} - b(l+\frac{1}{2})e^{-\beta/2}}.
\frac{ e^{\beta l+\frac{1}{2}} - b(l+\frac{1}{2}) }{ e^{\beta(l+1)} - b(l+1)e^{-\beta/2}}\nonumber\\
&&=
\frac{ e^{\beta l} - bl }{ e^{\beta(l+1)} - b(l+1)e^{-\beta}}.
\end{eqnarray}

This equality cannot be satisfied due to the factor of $e^{-\beta/2}$ on the bottom of the recursion relation.
As such, the discrete Penna model does not exhibit scale invariance. 
In the limit of a vanishing mutation rate: $e^{-\beta}$ approaches unity, 
the differences between scaled and unscaled models vanish and the discrete model will become scale invariant.
Fig.\ \ref{figScalingTrio} confirms that the scaled results of the discrete model do approach a limiting `master curve'.
This limit is the same as that which gives the continuous model, so we expect to find the continuous model
to be scale invariant. Note, only comparable points have been plotted and
distributions have been normalized so $\sum_l a^{-1} n(l) = 1$ where $a$ is the
scale factor. 

\subsection{The continuous Penna model}

As in the discrete Penna model, we identify a population by the largest value of $l_\max$ it can sustain. This
value is no longer constrained to be an integer and can be simply expressed as $l_\max=\frac{1}{\beta}$. 
Rescaling of a continuous Penna model is carried out in much the same way as in the discrete case: $l_\max$ is
divided by a scale factor $a$ and the new model has a correspondingly reduced maximum lifespan. In the continuous
model time is not broken into distinct timesteps, but is treated as a continuum: as a result 
rescaling will not alter
the number of timesteps in one unit of time. 
If a continuous model is to be invariant
under rescaling by a factor $a$, through similar reasoning as in the discrete case,
\begin{equation}
\frac{n_{a\lmax}(al+ax)}{n_{a\lmax}(al)}
=\frac{n_{\lmax}(l+x)}{n_{\lmax}(l)}.
\end{equation}
Upon substitution of the steady-state relation for continuous Penna model populations, this is satisfied by
\begin{eqnarray}
&&\frac{l+x}{l}\frac{e^{\beta l}-bl}{e^{\beta(l+x)}-b(l+x)}\nonumber\\
&&\times\exp\left[{\int_{l}^{l+x}\frac{\beta bl'}{bl'-e^{\beta l'}}\d l'}\right]\nonumber\\
&=&\frac{al+ax}{al}\frac{e^{\beta' al}-b'al}{e^{\beta'(al+ax)}-b'(al+ax)}\nonumber\\
&&\times\exp\left[{\int_{al}^{al+ax}\frac{\beta'b'al'}{b'al'-e^{\beta' al'}}\d al'}\right]
\end{eqnarray}
The mutation rate and birth rate in the rescaled model are labelled $\beta'$ and $b'$. If, by rescaling the model by 
a factor $a$, the maximum lifespans are to be the same then the mutation and birth rates must be 
related by: $\beta'=\beta/a$, 
$b'=b/a$. After this substitution the continuous model is clearly scale invaraint as both sides of the equation give
\begin{eqnarray}
\frac{l+x}{l}\frac{e^{\beta l}-bl}{e^{\beta(l+x)}-b(l+x)}\exp\left[{\int_{l}^{l+x}\frac{\beta bl'}
{bl'-e^{\beta l'}}\d l'}\right].
\end{eqnarray}

It has been shown that in the limit of a vanishing mutation rate coupled with an infinite maximum lifespan, 
the discrete model becomes a continuous one. In other words, for a vanishing mutation rate, the discrete model becomes
scale invariant. As the limits of a vanishing mutation rate and maximum genetic lifespan tending to infinity are 
equivalent, approximate scale invariance becomes more realistic for discrete Penna models of increasingly large $\lmax$. 

\section{Mortality rates}

Early age Penna mortality rates display the exponential growth predicted by the Gompertz law.
Using our analytical solution to the simple Penna model we evaluate the growth exponent $\gamma$
where the mortality rate at age $x$ is proportional to $e^{\gamma x}$. Evaluation of the Gompertz growth
rate in terms of Penna model parameters will facilitate the fitting of Penna parameters to real world data.
Throughout we assume that any model has adopted the maximum genetic lifespan allowed by its mutation rate.

\begin{figure}
\begin{center}
 \includegraphics[width=3in]{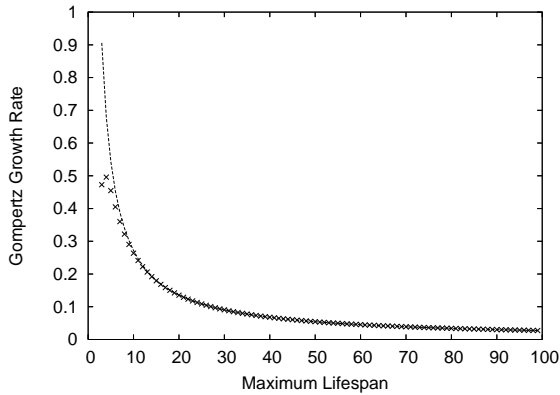}
 \caption{\label{figGompertz}The exponential coefficient of Gompertz growth in mortality rate estimated
from early age mortality rates ($\times$) is plotted against the maximum lifespan 
the population can sustain ($\lmax$). The dashed line gives the birth rate at each value of $\lmax$.}
\end{center}
\end{figure}

Recall that for the simple discrete Penna model the mortality rate is given by 
\begin{eqnarray}
{M}(x)=\frac{n(x)/x}
{\sum_{l=0}^\infty n(l)/l}.
\end{eqnarray}
Using the steady state recursion relation from the simple Penna model, the ratio between successive mortality
rates can be evaluated analytically
\begin{eqnarray}
\frac{M(x+1)}{M(x)}&=&\frac{e^{\beta x}-bx}{e^{\beta (x+1)}-b(x+1)e^\beta}\nonumber\\
&\times&\left[\frac{n(x)/x}{\sum_{l=x+1}^\infty n(l)/l} + 1\right].\label{mortalityRatio}
\end{eqnarray}
To usefully exploit this expression, we consider the limit of small $x$, and small $\beta$ where 
the Penna model becomes scale invariant; numerical evaluation of the summation term
and predicted scaling behaviour can be used to simplify equation (\ref{mortalityRatio}).
Numerically, we find for $x \ll l_{max}$
\begin{equation}
  \frac{n(x)/x}{\sum_{l=x+1}^\infty n(l)/l}\simeq \frac{1}{l_{max}}.
\label{numericalMortalityResult}
\end{equation}
Crucially, if the Penna model exhibits universality as discussed earlier, this result remains valid for
{\it all} values of $\lmax$. Therefore, noting the continuous Penna
model result $\lmax=1/\beta$ and $b=\beta e$, in the regime of small $x
\ll \lmax$, a first order expansion of equation (\ref{mortalityRatio})
leads to
\begin{eqnarray}
\frac{M(x+1)}{M(x)}&\approx&e^{b}
\end{eqnarray}
which then implies
\begin{equation}
M(x) \propto e^{b x},
\end{equation}
namely the Gompertz law, which states that the mortality rate
increases exponentially at early ages. Furthermore, it predicts that
the exponential coefficient of the Gompertz growth rate is given
by $b$, the birth rate. In Fig.~\ref{figGompertz}, we compare this
birth rate with the exponential Gompertz coefficients, extracted by taking the difference between the logs of mortality
rates at ages $x=2$ and $x=1$ for each population.

Our approximation depends on $x \ll \lmax$, therefore, deviation from
Gompertz behaviour at later ages (large $x$) is expected as the numerical approximation, 
equation (\ref{numericalMortalityResult}), breaks down as $x$
increases. Similarly, as shown in Fig.~\ref{figGompertz}, for small
values of $\lmax$ this approximation works less well but for
larger values of $\lmax$ it becomes increasingly accurate.

\section{Conclusion}
We have shown by means of exact analytic solution that, in the asexual
Penna model, a universal scale invariance emerges as we increase the
number of genome bits/sites, with the invariance becoming exact in the
limit of the continuum model. In addition, we have built on this
result and shown that scale invariance may be employed to derive an 
analytical expression for the Gompertz law of mortality, which has been generally
regarded as empirical. 

\acknowledgements{
The authors would like to thank Krzysztof Malarz for helpful discussion.
This work has been financially supported by the Schiff foundation 
and the EPSRC under grant number GR/TR11777/01.

This paper is dedicated to Freddie Mao, a recent addition to the Mao family.
}

\end{document}